\documentclass[8.5pt,twoside,twocolumn]{article}
\usepackage[super,sort&compress,comma]{natbib} 
\usepackage{mhchem}
\usepackage{times,mathptmx}
\usepackage{sectsty}
\usepackage{balance} 

\usepackage{graphicx} 
\usepackage{lastpage}
\usepackage[format=plain,justification=raggedright,singlelinecheck=false,font=small,labelfont=bf,labelsep=space]{caption} 
\usepackage{fancyhdr}
\usepackage{fixltx2e}
\usepackage{pslatex}
\usepackage{graphicx}

\usepackage{color}

\begin{document}

\thispagestyle{plain}
\renewcommand{\thefootnote}{\fnsymbol{footnote}}
\renewcommand\footnoterule{\vspace*{1pt}
\hrule width 3.4in height 0.4pt \vspace*{5pt}} 
\setcounter{secnumdepth}{5}

\makeatletter 
\def\subsubsection{\@startsection{subsubsection}{3}{10pt}{-1.25ex plus -1ex minus -.1ex}{0ex plus 0ex}{\normalsize\bf}} 
\def\paragraph{\@startsection{paragraph}{4}{10pt}{-1.25ex plus -1ex minus -.1ex}{0ex plus 0ex}{\normalsize\textit}} 
\renewcommand\@biblabel[1]{#1}            
\renewcommand\@makefntext[1]%
{\noindent\makebox[0pt][r]{\@thefnmark\,}#1}
\makeatother 
\renewcommand{\figurename}{\small{Fig.}~}
\sectionfont{\large}
\subsectionfont{\normalsize} 

\fancyfoot{}
\fancyhead{}
\renewcommand{\headrulewidth}{1pt} 
\renewcommand{\footrulewidth}{1pt}
\setlength{\arrayrulewidth}{1pt}
\setlength{\columnsep}{6.5mm}
\setlength\bibsep{1pt}

\twocolumn[
  \begin{@twocolumnfalse}

\noindent\LARGE{\textbf{Non-Grotthuss Proton Diffusion Mechanism in Tungsten Oxide Dihydrate from First-Principles Calculations$^\dag$}}

\vspace{0.6cm}

\noindent\large{Hao Lin, Fei Zhou,\textit{$^{\ddag}$} Chi-Ping Liu, and
Vidvuds Ozoli\c{n}\v{s}$^{\ast}$}\vspace{0.5cm}

\noindent\textit{\small{\textbf{Received Xth XXXXXXXXXX 20XX, Accepted Xth XXXXXXXXX 20XX\newline
First published on the web Xth XXXXXXXXXX 200X}}}

\noindent \textbf{\small{DOI: 10.1039/C4TA02465F}}
\vspace{0.6cm}

\noindent \normalsize{Fast proton conduction mechanism is of key importance for achieving high performance in fuel cell membranes, batteries, supercapacitors, and electrochromic materials. Enhanced proton diffusion is often observed in hydrated materials where it is thought to occur via the famous Grotthuss mechanism through pathways formed by structural water. Using first-principles density-functional theory calculations, we demonstrate that proton diffusion in tungsten oxide dihydrate (\ce{WO3.2H2O}), a known good proton conductor, takes place within the layers of corner-sharing \ce{WO6} octahedra without direct involvement of structural water. The calculated proton migration barrier in \ce{WO3.2H2O} (0.42 eV) is in good agreement with the experimental value inferred from the temperature dependence of conductivity (0.36 eV). The preferred proton diffusion path in \ce{WO3.2H2O} is essentially the same as in $\gamma$-\ce{WO3}, and we find an activation energy of 0.35 eV for the latter, which agrees well with the experimental values. In contrast to the small intercalation voltages calculated for \ce{WO3} and \ce{WO3.2H2O}, we find that proton absorption in the monohydrate \ce{WO3.H2O} is energetically highly favorable, corresponding to voltages in excess of 1 eV in the dilute limit. However, strong proton-proton repulsion limits the equilibrium H content at zero voltage. We find a fast one-dimensional diffusion channel in \ce{WO3.H2O} with an activation energy of only 0.07 eV at dilute proton concentrations, but much higher barriers are expected at near-equilibrium concentrations due to strong repulsive interactions with other protons. Our results illustrate that low proton diffusion barriers and low insertion voltages both contribute to fast proton transport in bulk \ce{WO3.2H2O} and $\gamma$-\ce{WO3}.}
\vspace{0.5cm}
 \end{@twocolumnfalse}
 ]

\section{Introduction}
\label{sec:intro}

\footnotetext{\dag~Electronic Supplementary Information (ESI) available: Band structures, electronic density of states and proton diffusion movies. See DOI: 10.1039/C4TA02465F}

\footnotetext{\textit{~Department of Materials Science and Engineering, University of California, Los Angeles, P.O. Box 951595, Los Angeles, California 90095-1595, USA; E-mail: vidvuds@ucla.edu}}


\footnotetext{\ddag~Present address: Condensed Matter and Materials Division,
Lawrence Livermore National Laboratory, Livermore, California 94550, USA}

Fundamental understanding of the mechanisms of proton conduction is crucial for the development of fuel cell membranes, batteries, supercapacitors and electrochromics. \cite{di1978electrochromism,xie2012fast,sugimoto2006charge} It is commonly believed that proton diffusion in hydrous materials occurs via a Grotthuss type mechanism facilitated by water molecules.\cite{Grotthuss1806proton} For instance, fast protonic transport in hydrous ruthenia (\ce{RuO2.$x$H2O}) has been attributed to the existence of structural water at grain boundaries. \cite{sugimoto2006charge} The appearance of a strongly rate-dependent contribution to the charging capacity of ruthenia at very slow rates suggests that proton diffusion in bulk \ce{RuO2} is kinetically hindered, \cite{sugimoto2004evaluation} a conclusion which is supported by a high migration barrier found in first-principles calculations. \cite{ozolins2013ruthenia} Recently, tungsten oxide dihydrate (\ce{WO3.2H2O}) was observed to have a relatively high proton conductivity (7$\times$$10^{-3}$ S/cm at 423 K) and low proton activation energies for bulk (0.36 eV) and surface (0.15 eV) diffusion, \cite{li2000proton} demonstrating its potential for use as a proton conductor at low and medium temperatures (273--423 K). The presence of layered water in the crystal structure of \ce{WO3.2H2O} again seemingly suggests that the Grotthuss mechanism contributes to proton diffusion through this compound, but our knowledge of the actual proton diffusion kinetics in \ce{WO3. 2H2O} is limited. In particular, it remains an open question whether proton transport in \ce{WO3.2H2O} is mediated by the structural water layer in this compound. A related question is whether the much higher bulk proton conductivity of \ce{WO3.2H2O} in comparison with the monohydrate, \ce{WO3.H2O}, can be explained by the absence of hydrogen-bonded layers of structural water in the latter.

This paper reports a comparative first-principles study of proton intercalation and transport in a series of structurally related solids ($\gamma$-\ce{WO3}, \ce{WO3.H2O}, and \ce{WO3.2H2O}) that incorporate corner-linked networks of \ce{WO6} octahedra. Using density-functional theory (DFT) calculations, we investigate the crystal structures, bonding, electronic properties, proton intercalation energetics, and energy barriers for proton diffusion. We find that intercalated hydrogen donates charge to the empty orbitals in the tungsten oxide layer. Surprisingly, our results show that proton diffusion in \ce{WO3.2H2O} occurs through the layers of corner-sharing \ce{WO6} octahedra and does not involve the hydrogen-bonded \ce{H2O} network in the structural water layer. This is explained by the fact that the preferred adsorption site for the intercalated proton is at one of the bridging oxygen ions where it forms a bond with a nonbonding O $2p$ orbital. Diffusion is facilitated by a low energy cost for a concerted rotation of the interlinked \ce{WO6} octahedra that is involved in "swinging" the adsorbed proton around the bridging oxygen ion. Our calculations show that this configuration is approximately 0.3 eV more favorable than the formation of a hydronium ion (\ce{H3O+}) in the water layer. Since the anhydrous $\gamma$-phase \ce{WO3} also contains a corner-linked network of \ce{WO6} octahedra, we find that proton diffusion in this material follows the same mechanism and has a similar activation energy. A more complex behavior is predicted for the monohydrate \ce{WO3.H2O}. At dilute concentrations, protons adsorb at the terminating oxygen site where they can diffuse along one-dimensional zig-zag pathways running in the [100] direction between the octahedral layers. The calculated barrier of proton migration along the one-dimensional pathways is low (0.07 eV). In a marked difference from  $\gamma$-\ce{WO3} and \ce{WO3.2H2O}, hydrogen absorption in  \ce{WO3.H2O} is energetically highly favorable (by more than 1 eV/H) and the dilute limit is not experimentally relevant. We calculate that at non-dilute concentrations the majority of protons adsorb within the octahedral layers and strong proton-proton repulsion limits the equilibrium zero-voltage hydrogen concentration in \ce{H$_{y}$WO3.H2O} to $y$ values in the 10--20\% range. To explain why only surface diffusion contribution is observed for \ce{WO3.H2O},\cite{li2000proton} we hypothesize that the one-dimensional diffusion channels are blocked due to repulsive interactions with the protons in the octahedral layers. We also suggest that the bulk diffusion mechanism predicted for \ce{WO3} and \ce{WO3.2H2O} is not operational in \ce{WO3.H2O} at non-dilute concentrations, because adsorbed protons cause a stiffening of the octahedral layer with respect to the rotations of the \ce{WO6} octahedra, inhibiting a key step in the proton diffusion mechanism.

\section{Methods}
\label{sec:methods}

The Perdew-Burke-Ernzerhof (PBE) \cite{perdew1996generalized} exchange correlation functional and the projector augmented wave (PAW) method \cite{blochl1994projector} as implemented in the Vienna {\it Ab Initio} Simulation Package (VASP)\cite{kresse1996efficiency} were used in all our calculations. We used bulk supercells of $\gamma$-\ce{WO3} (\ce{WO3.H2O}/\ce{WO3.2H2O}) containing 32 (16/16) formula units. In all cases, plane wave basis sets with an energy cutoff of 875 eV were used to expand the electronic wave functions, and 2$\times$2$\times$2 Monkhorst-Pack \cite{monkhorst1976special} {\bf k} point meshes were used to sample the Brillouin zone. Convergence tests showed that  with these settings the total energies were converged to within 2 meV per formula unit, compared to calculations with a 10$\times$10$\times$10 Monkhorst-Pack {\bf k} point mesh. Atomic coordinates were fully relaxed until all forces were below 0.02 eV/{\AA} and cell parameters were relaxed until components of the stress tensor were below 0.4 kbar. Our tests for tungsten oxide dihydrate showed that spin-orbit coupling (SOC) effects increase the calculated proton intercalation voltage by only 0.03 eV, and hence SOC is neglected in the results reported here. Proton diffusion pathways between two locally stable proton sites are calculated using the nudged elastic band (NEB) method\cite{jonsson1998nudged} with at least five intermediate images. The transition state obtained by NEB was further refined using the climbing nudged elastic band (cNEB) method. \cite{henkelman2000climbing} Activation barriers presented here include quantum tunneling corrections calculated at 298 K using the formalism of Fermann and Auerbach.\cite{fermann2000modeling}

In electrochemical experiments, proton intercalation is accompanied by an electron insertion. This process is modeled computationally as insertion of neutral hydrogen atoms.  Using the reversible hydrogen electrode as a reference, the voltage of proton insertion in the dilute limit is calculated as\cite{ozolins2013ruthenia}
\begin{equation}
\label{eq:voltage}
V= \frac{\Delta G}{e},
\end{equation}
where  $ \Delta G$ is the free energy of proton intercalation in bulk \ce{WO3.$x$H2O} ($x=0,1,2$).  $ \Delta G$ can be calculated from the following expression:
\begin{equation}
\begin{split}
\Delta G = &G[\text{H}_{m+1} \ce{(WO3.$x$H2O)}_n]- G[\text{H}_m\ce{(WO3.$x$H2O)}_n] \\
                 &-\frac{1}{2} G^\circ [\ce{H2}]
\end{split}
\end{equation}
where $G[\text{H}_m\ce{(WO3.$x$H2O)}_n]$ is the total free energy of bulk supercell of \ce{WO3.$x$H2O} with $m$ additional hydrogen atoms, $G[\text{H}_{m+1}\ce{(WO3.$x$H2O)}_n]$ is the total free energy of bulk \ce{WO3.$x$H2O} with $m+1$ additional hydrogen atoms, and $ G^\circ $[\ce{H2}] is the standard state free energy of hydrogen gas at atmospheric pressure and $T=298$~K.\cite{ozolins2013ruthenia}

\section{Results}
\label{sec:results}

\subsection{Structural properties}
\label{sec:structure}

\begin{figure*}[h]
\centering
  \includegraphics[height=10cm]{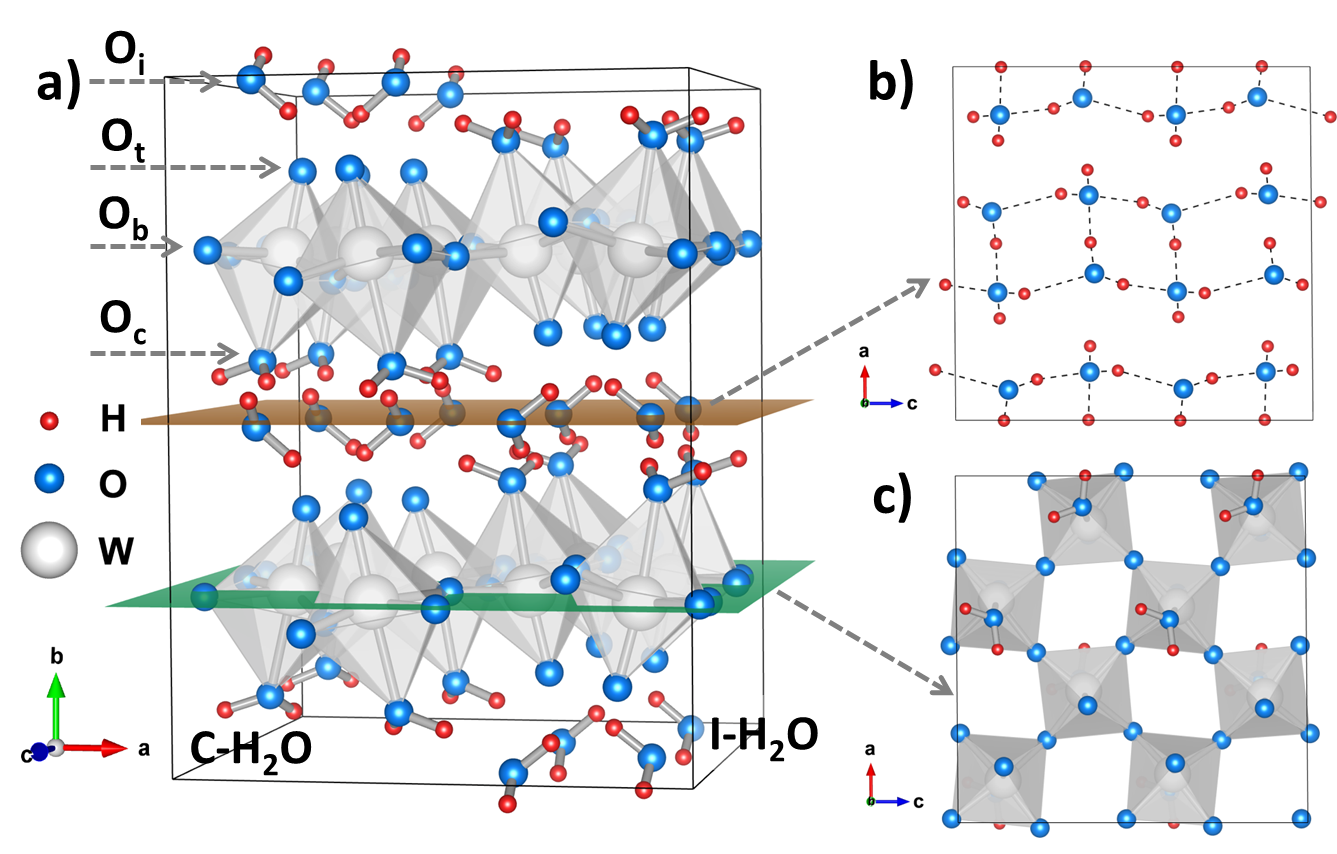}
  \caption{a) Crystal structure of tungsten oxide dihydrate, where O\textsubscript{c}, O\textsubscript{b}, O\textsubscript{t} and O\textsubscript{i} represent coordinated water oxygen, bridging oxygen, terminating oxygen and oxygen in interlayer water respectively, and C-\ce{H2O}, I-\ce{H2O} denote coordinated water and interlayer water, b) hydrogen-bonded network, c) a tungsten-oxygen octahedral layer. Protons are shown in red, oxygen ions in blue, tungsten ions in white and \ce{WO6} octahedra in grey.
 }
  \label{fig:dihydrate}
\end{figure*}

The crystal structure of tungsten oxide dihydrate (shown in Fig.~\ref{fig:dihydrate}a) belongs to the monoclinic {\it $P2_{1}/n$} space group.  This structural framework consists of a connected network of corner-sharing \ce{WO6} octahedra (shown in grey in Fig.~\ref{fig:dihydrate}c) and two types of water molecules. The first type, called "coordinated water", shares its oxygen with the tungsten ion, while the other, referred to as "interlayer water", is located within the (010) plane between the layers of \ce{WO6} octahedra (see Fig.~\ref{fig:dihydrate}a and \ref{fig:dihydrate}b). As marked in Fig.~\ref{fig:dihydrate}a, there are four types of oxygen sites in \ce{WO3.2H2O}: coordinated water oxygen (O\textsubscript{c}), bridging oxygen (O\textsubscript{b}), terminating oxygen (O\textsubscript{t}), and interlayer water oxygen (O\textsubscript{i}). Within the distorted octahedra, the relaxed W-O\textsubscript{b} bond lengths range from 1.86 to 2.03 {\AA}, and the distance between the W and O\textsubscript{t} ions is 1.75 {\AA}, much shorter than the length of the W-O\textsubscript{c} bond (2.31 {\AA}). The variation in W-O bond lengths is due to different types of bonding: (1) an O\textsubscript{t} ion is only bonded to a W ion, (2) each O\textsubscript{b} ion is bonded to two W ions within octahedral planes, and (3) weak W-O\textsubscript{c} bonding because of  strong O\textsubscript{c}H bonding in the $sp^3$ hybridization. A more detailed analysis of the electronic structure and bonding in \ce{WO3.2H2O} is given in Section~\ref{sec:elstruc}.

Rietveld refinement results for \ce{WO3.2H2O} do not contain the coordinates of hydrogen atoms because the XRD patterns of \ce{WO3.2H2O} were determined by Cu-K$\alpha$ radiation (1.54056 {\AA}), which has a longer wavelength than typical O-H distances. \cite{li2000proton} Since \ce{WO3.2H2O} and \ce{MoO3.2H2O} share the same space group and exhibit identical arrangements of the metal and oxygen ions, we obtained the structure of \ce{WO3.2H2O} by taking the hydrogen positions from the crystal structure of \ce{MoO3.2H2O} as input and relaxing all degrees of freedom.  The calculated $a$, $b$ and $c$ lattice parameters are 10.57, 14.12, and 10.67 {\AA}, respectively, slightly larger than the experimental values ($a=10.48$, $b=13.97$, and $c=10.62$~{\AA}), while the relaxed value of the monoclinic angle $\beta = 90.48 ^{\circ} $ is slightly smaller than that in the experiment ($91.59 ^{\circ}$). The differences of crystal parameters between our DFT calculations and the XRD data are below 3\% and can be attributed to the approximate nature of the PBE exchange-correlation functional, as has been well documented in the computational chemistry community. \cite{fuchs1998pseudopotential,fuchs2002cohesive,fischer2006predicting} It is also found that the discrepancy of the lattice parameter in the {\it b} direction is slightly larger than in the other two directions. This can be attributed to the inappropriate description of the van der Waals force between the interlayer water molecules and the \ce{WO6} octahedra in the {\it b} direction arising from the use of the PBE exchange correlation functional in our calculations.\cite{perdew1996generalized}

To further verify that the generated crystal structure is reasonable, we calculated the reaction enthalpy of the dehydration reaction $\ce{WO3.2H2O} \longrightarrow \ce{WO3.H2O} + \ce{H2O}(l)$, defined by:
\begin{equation}
\label{eq:enthalpy}
\Delta H =  H[\ce{WO3.H2O}] + H^\circ[\ce{H2O}(l)] - H[\ce{WO3.2H2O}]
\end{equation}
where $H[\ce{WO3.H2O}]$, $H[\ce{WO3.2H2O}]$ and $H[\ce{H2O}(l)]$ are the enthalpies of crystalline \ce{WO3.H2O} and \ce{WO3.2H2O}, and liquid \ce{H2O}, respectively. The enthalpies of \ce{WO3.H2O} and \ce{WO3.2H2O} can be well approximated by the DFT total energies whereas accurate enthalpy value for liquid \ce{H2O} is difficult to obtain from DFT directly. Here, we approximated $H^\circ [\ce{H2O}(l)]$ by using the reaction enthalpy $\Delta H^\circ _{r} $ (3 kJ/mol) of another dehydration reaction, $\ce{WO3.H2O} \longrightarrow \ce{WO3} + \ce{H2O}(l)$, under standard conditions, \cite{chase1998nist} and the DFT total energies of \ce{WO3.H2O} and \ce{WO3}:
\begin{equation}
H^\circ[\ce{H2O}(l)] \approx E[\ce{WO3.H2O}] - E[\ce{WO3}]+ \Delta H^\circ _{r}.
\end{equation}
The reaction enthalpy obtained from Eq.~(\ref{eq:enthalpy}) is 36 kJ/mol, while the experimentally measured value is 40 kJ/mol. \cite{li2000proton} This good agreement is an indirect justification of the validity of our computational methodology.

\begin{figure}[!htbp]
\centering
\includegraphics[width=7.5cm]{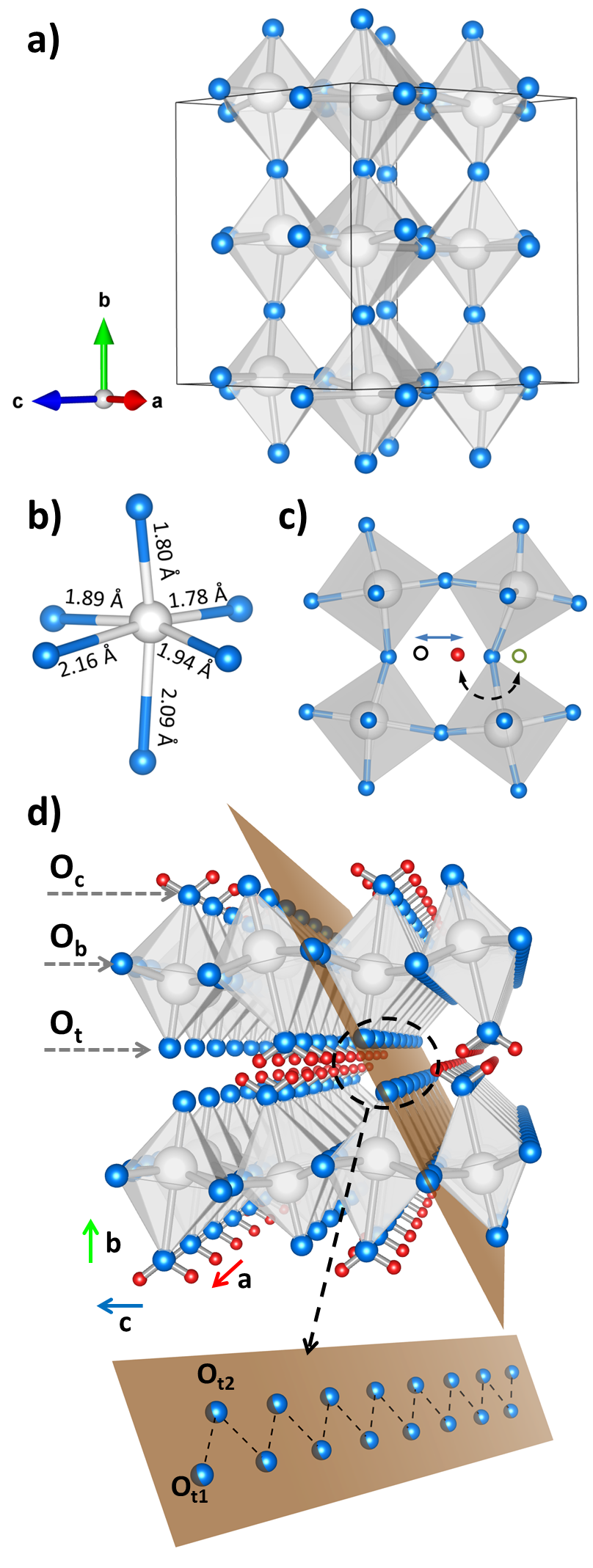}
\caption{(a) Crystal structure of  anhydrous monoclinic $\gamma$-phase of tungsten oxide, $\gamma$-\ce{WO3}. (b) An octahedron in $\gamma$-\ce{WO3}. (c) Proton diffusion in $\gamma$-\ce{WO3}. (d) Crystal structure of monoclinic tungsten oxide monohydrate, \ce{WO3.H2O}. 
}
\label{fig:monohydrate}
\end{figure}

With the same space group ({\it $P2_{1}/n$}) as tungsten oxide dihydrate, the room-temperature monoclinic phase of tungsten oxide ($\gamma$-\ce{WO3}) is characterized by a three-dimensional (3D) network of corner-sharing \ce{WO6} octahedra that are aligned along the $b$ axis, as shown in Fig.~\ref{fig:monohydrate}a. W ions are off-center in the octahedra, which results in six different W-O bond lengths (see Fig.~\ref{fig:monohydrate}b). The W-O bond lengths in the $a$ direction are approximately the same (1.89 and 1.94 {\AA}), while in the $b$ and $c$ directions the W-O distances vary from 1.78 to 2.16 {\AA}. These distortions have been attributed to a second order Jahn-Teller effect. \cite{Goodenough1971145}

Tungsten oxide monohydrate differs from the dihydrate by the absence of the hydrogen-bonded interstitial water layer, but it contains a similar network of corner-sharing \ce{WO6} octahedra with three types of oxygen ions (see Fig.~\ref{fig:monohydrate}d): bridging oxygen ions (O\textsubscript{b}) connecting the octahedra within the layer, each shared by two W$^{6+}$ ions, and terminating and coordinated oxygen ions (O\textsubscript{t} and O\textsubscript{c}, respectively) on the opposite tips of each \ce{WO6} octahedron; O\textsubscript{c} is bonded to two hydrogen ions, forming a water molecule. We point out that the O\textsubscript{t} oxygen ions are arranged in one-dimensional (1D) rows running along the $a$ direction and, as will be shown below, these rows act as easy diffusion channels in the dilute limit.

\subsection{Electronic structure}
\label{sec:elstruc}

\begin{figure*}[h]
\centering
\includegraphics[width=17cm]{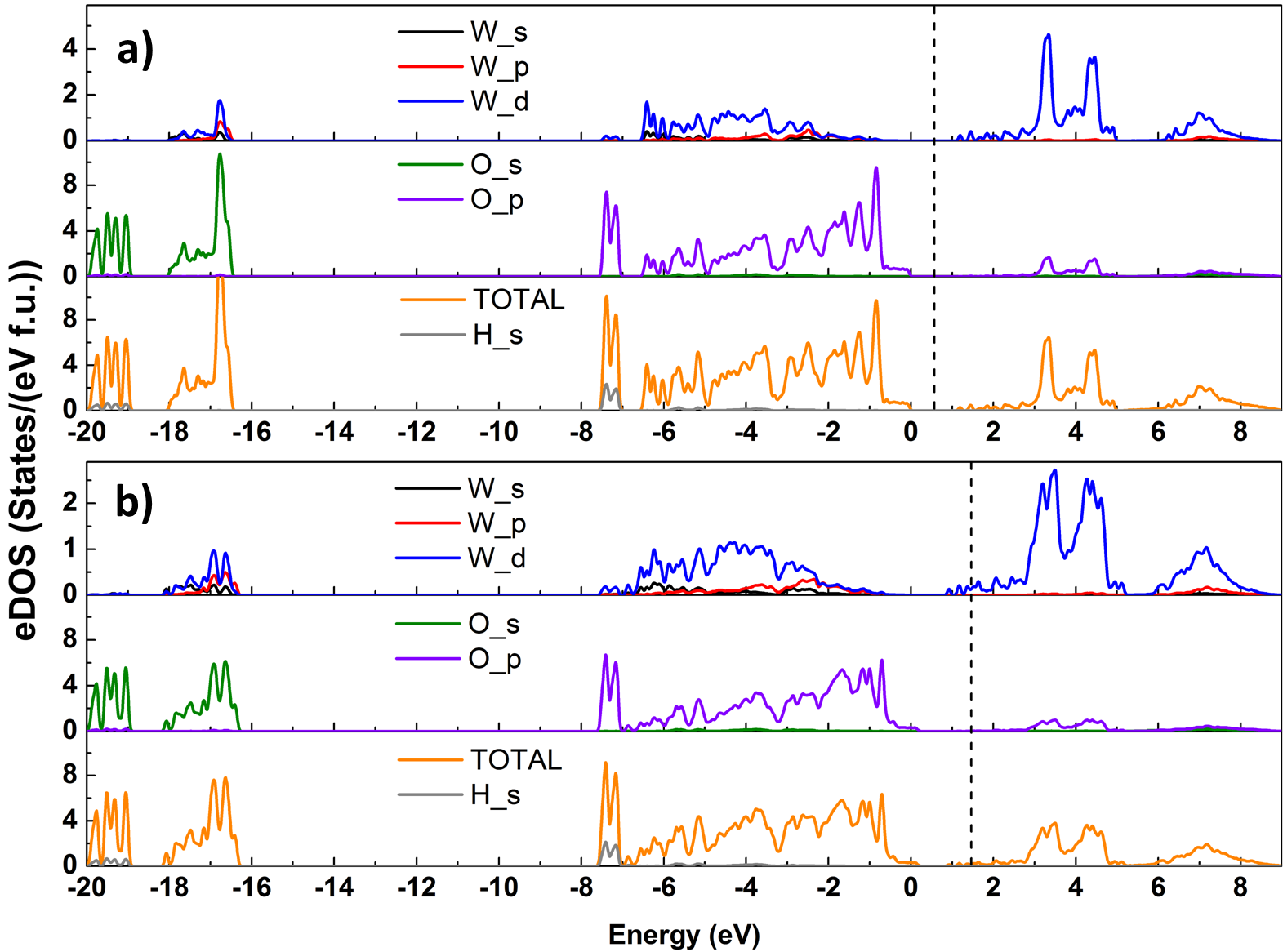}
\caption{Electronic density of states (eDOS) of (a) pure tungsten oxide dihydrate and (b) proton-intercalated tungsten oxide dihydrate. The fermi levels are indicated by dashed lines.}
\label{fig:eDOS}
\end{figure*}

Tungsten oxide dihydrate is predicted to be a direct-gap semiconductor with a calculated PBE band gap of 1.16 eV, as shown in Fig.~S1 in the Electronic Supplementary Information (ESI).\dag For comparison, the calculated values of band gaps for \ce{WO3.H2O} and $\gamma$-\ce{WO3} are 0.85 and 1.34 eV (see Fig.~S2 and S3 in ESI\dag), significantly smaller than the experimental values of 2.17 eV and 2.6 eV, respectively. \cite{wang2009ag,koffyberg1979interband, salje1974new} To the best of our knowledge, experimental values of the band gap of \ce{WO3.2H2O} have not been reported. Since the color of dihydrate samples is yellow, \cite{freedman1959tungstic} the band gap is expected to be larger than 2 eV. It is well understood that nonlocal exchange correlation (xc) functionals are needed to achieve satisfactory agreement between the calculated and experimental band gaps in \ce{WO3}.\cite{ping2013optical,wang2011electronic} However, intercalation voltages and diffusion barriers, which are of main interest here, are expected to be less sensitive to nonlocal xc corrections if the additional electrons occupy delocalized $d$ bands.\cite{zhou2004ldapu} Good agreement between the calculated and experimentally measured intercalation voltages obtained below {\it a posteriori\/} justifies our use of the semi-local PBE functional. Finally, we note that the electronic bands of both hydrates are flat in the direction perpendicular to the octahedral layers (see Fig.~S1 and S2 in ESI\dag), indicating that there is practically no wave function overlap across the gap between the layers.

The calculated electronic density of states (eDOS) of \ce{WO3.2H2O} is shown in Fig.~\ref{fig:eDOS}a, and the atom- and angular momentum-decomposed eDOS can be found in Fig.~S4 in ESI.\dag The energy bands can be divided into three groups: a) states between $-20$ and $-16$ eV below the Fermi level, mainly originating from the oxygen $2s$ orbitals, b) states between $-8$ and $0$ eV, chiefly formed by the oxygen $2p$ orbitals, and c) bands above the Fermi level of predominantly tungsten $5d$ character. Specifically, the deep-lying valence bands from $-20$ to $-19$ eV are due to the $2s$ orbitals of the O\textsubscript{i}/O\textsubscript{c} ions.  Both the O\textsubscript{t} and O\textsubscript{b} $2s$ orbitals are overlapped by W $5d$, $6s$, $6p$ orbitals in the octahedral field, yielding $\sigma$ bonded states in the region from $-18$ to $-16$ eV. The sharp peaks in the range from $-8$ to $-7$ eV originate from $\sigma$ bonding hybrids between O\textsubscript{i}/O\textsubscript{c} $2p$ orbitals and H $1s$ orbitals. States between $-7$ and $-1.5$ eV are $\sigma$/$\pi$ bonding hybrids formed by hybridizing O $2p$ and W $5d$ orbitals, and the corresponding antibonding hybrids are in the conduction bands between $1$ and $9$ eV. We observe that the peaks in the $-1.5$ to $-0.5$ eV range are from the nonbonding $p$ orbitals of the O\textsubscript{b} ions, which point perpendicular to the tungsten-oxygen octahedral layers. They are shown to play an important role in the proton adsorption and diffusion mechanisms described in Section \ref{sec:diffusion}.

\subsection{Proton intercalation sites in dihydrate}
\label{sec:protons}

\begin{figure}
\centering
\includegraphics[width=8cm]{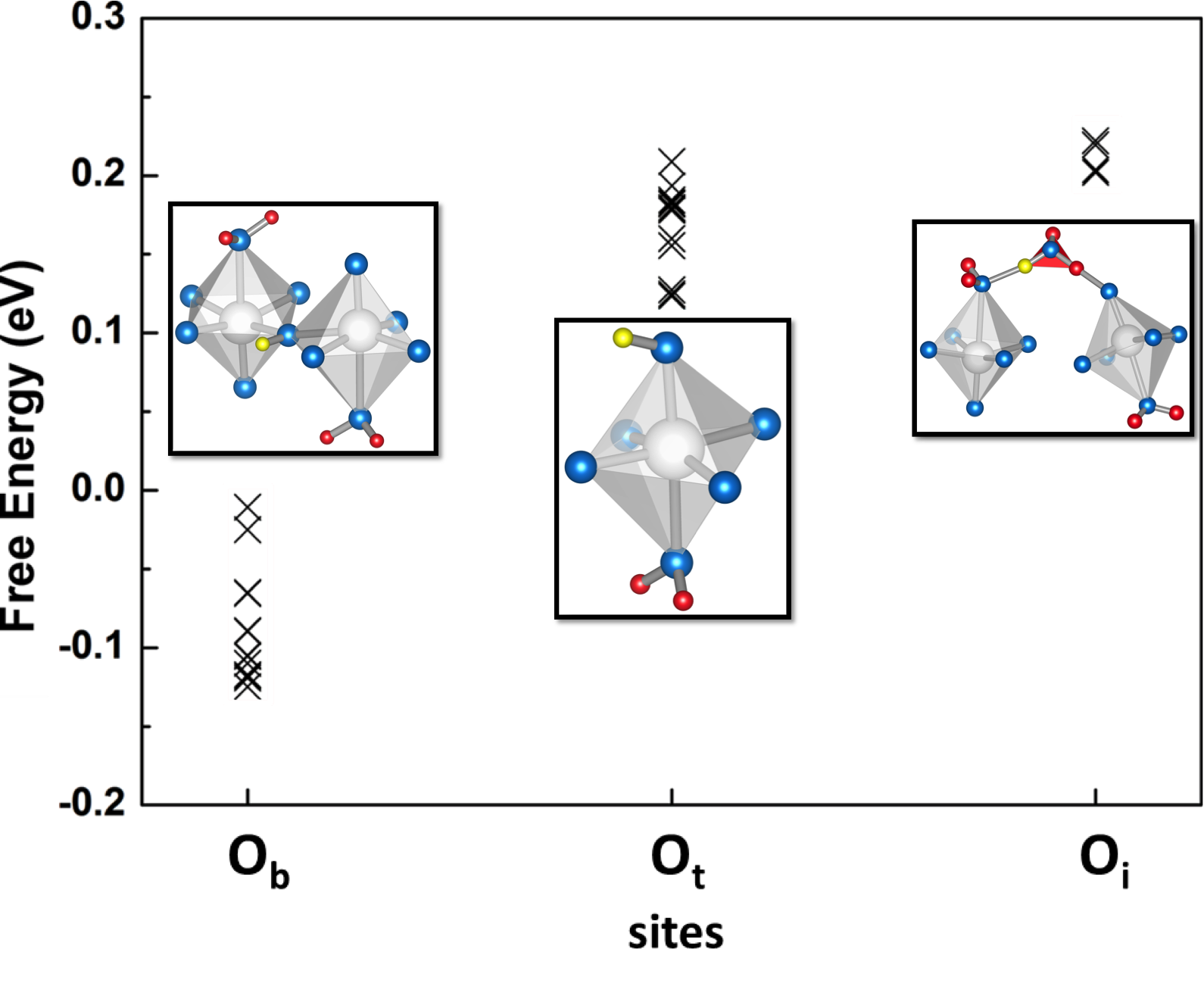}
\caption{Free energies of proton intercalation in \ce{WO3.2H2O}. Inserted protons are shown in yellow.}
\label{fig:protonE}
\end{figure}

\begin{figure}
\centering
\includegraphics[width=7.5cm]{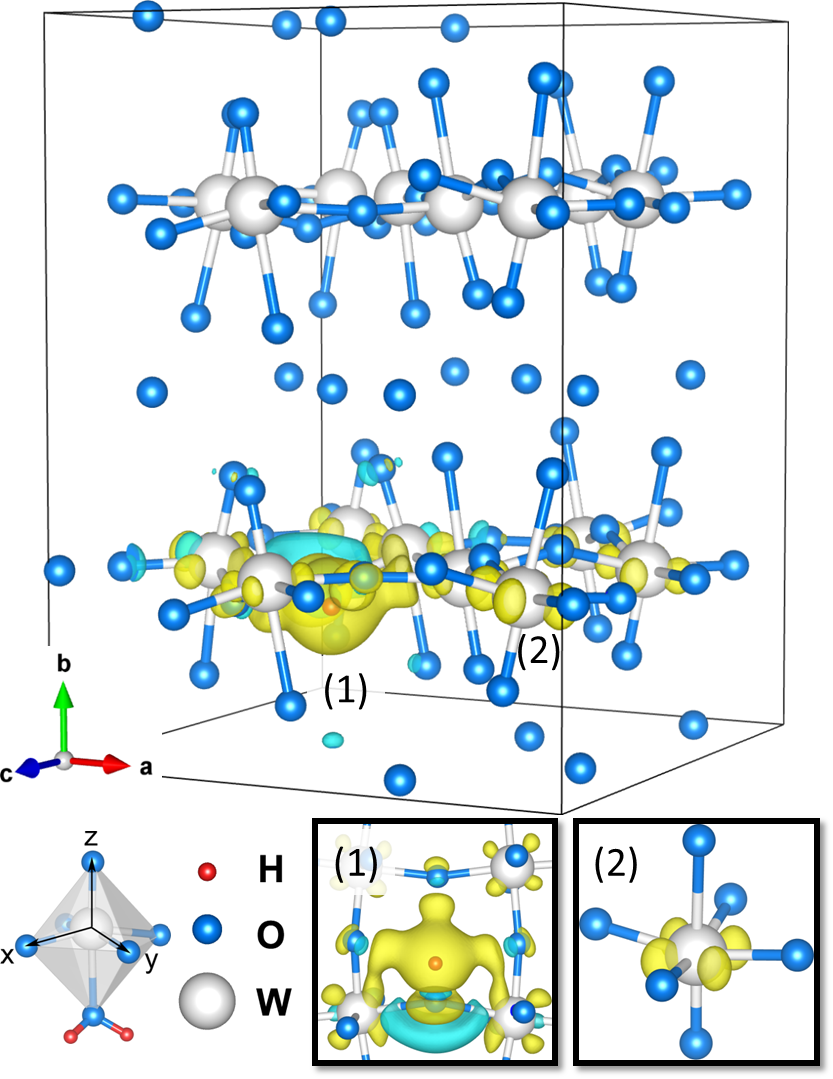}
\caption{Charge density distribution of an additional electron in proton-intercalated \ce{WO3.2H2O} relative to \ce{WO3.2H2O}. To facilitate the comparison, the ions in \ce{WO3.2H2O} are fixed to their positions in proton-intercalated \ce{WO3.2H2O}. For visual clarity, the hydrogen ions are not shown except the inserted proton (red). The electron accumulation region is shown in yellow while the depletion region is green. To describe the $d$ orbitals in the \ce{WO6} octahedra, a new coordinate system, rotated by 45 degrees around the $b$ axis, is introduced.}
\label{fig:chdens}
\end{figure}

The candidate intercalation sites for external protons in \ce{WO3.2H2O} were determined by constructing a regular real-space mesh with a spacing of 0.2 {\AA} and choosing those grid points that satisfy the following criteria: (1) the distance between the extra proton and at least one oxygen ion is between 0.8 and 1.8 {\AA}, (2) the distance between a tungsten ion and the additional proton is above 1.2 {\AA}, and (3) the distance between a native hydrogen ion and the intercalated proton is larger than 1.0 {\AA}. These criteria take into account the fact that hydrogen and tungsten ions are positively charged and therefore repel each other and that stable proton intercalation sites will be near the O$^{2-}$ anions.  For all sites satisfying these criteria, the structure was relaxed to the local energy minimum. Fig.~\ref{fig:protonE} shows the free energies of proton intercalation in \ce{WO3.2H2O} for different sites at 298 K. The spread of the energies for each type of oxygen site in Fig.~\ref{fig:protonE} is due to the intrinsic distortion of the \ce{WO6} octahedra and due to the different orientations of the two types of water molecules near the intercalation sites.  As discussed in Section~\ref{sec:structure} and Fig.~\ref{fig:dihydrate}, there are four types of oxygen sites in \ce{WO3.H2O}.  It was found that protons prefer the bridging oxygen sites, where the lowest free energy is -0.12 eV, corresponding to a proton insertion voltage of 0.12 V. For the other three types of oxygen sites, the free energies are positive, indicating that proton adsorption at these oxygen sites is thermodynamically unfavorable at 298 K. Specifically, a proton at the O\textsubscript{c} site is both thermodynamically unfavorable and kinetically unstable, because it repels the coordinated water oxygen while relaxing towards and bonding to either one of the terminating oxygen ions or to an oxygen ion in the interlayer water.

The proton-intercalated structure with the lowest total energy was used for further electronic structure analysis. After introducing a hydrogen atom into the system, tungsten oxide dihydrate becomes metallic (see Fig.~\ref{fig:eDOS}b). In comparison with the electronic density of states (eDOS) of pure \ce{WO3.2H2O}, the eDOS of proton-intercalated \ce{WO3.2H2O} shifts the Fermi level from the band gap to the bottom of the conduction band formed by empty $\pi^*$ bonding originating from W $5d$ and bridging oxygen $2p$ orbitals. This interpretation is further supported by visualizing the electronic charge density distribution of the extra electron (shown in Fig.~\ref{fig:chdens}). It can be seen that this electron is largely confined around the O-H group with some delocalization into the W 5${\it d}_{xy}$ orbitals within the same tungsten layer.  The extra electron charge density does not delocalize between the layers because there is no wave function overlap across the gap, in agreement with the existence of a flat band dispersion perpendicular to the layers (see Fig.~S1 in ESI\dag). 
The formation of small polarons is not expected in \ce{WO3.2H2O} because the structure of the \ce{WO6} octahedra is similar to that in crystalline \ce{WO3}, and polarons are not favored in the latter.\cite{zheng2011nanostructured, deepa2006case}

\subsection{Proton diffusion}
\label{sec:diffusion}

\subsubsection{WO\textsubscript{3}$\cdot$2H\textsubscript{2}O. }

\begin{figure}[!htbp]
\centering
\includegraphics[width=8cm]{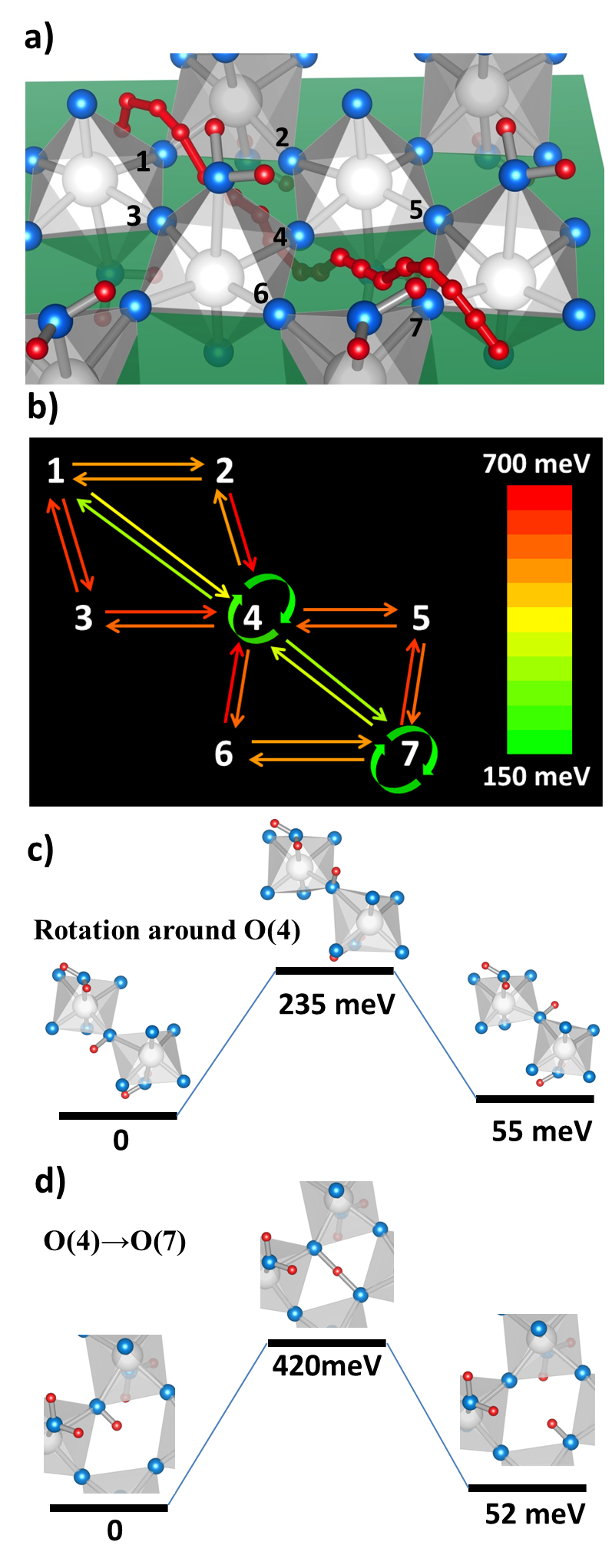}
\caption{Proton diffusion in dihydrate: a) diffusion path, b) activation barriers, and transition states of c) proton rotation around O(4) and d) proton hopping from O(4) to O(7).
}
\label{fig:diffusion}
\end{figure}

In water and in hydrates where water molecules form hydrogen bonds, proton diffusion can proceed via the Grotthuss mechanism. \cite{marx2006proton} In this mechanism, a proton is transferred from hydronium H$_3$O$^+$ to another water molecule via an intermediate formation of a Zundel cation H$_5$O$_2^+$, which consists of two water molecules sharing a proton. Unexpectedly, we do not observe this mechanism in \ce{WO3.2H2O}. There are two reasons for this behavior. First, a coordinated water molecule is not able to form a Zundel cation (\ce{H5O2+}) with another water molecule, because the O\textsubscript{c} ion is one of the ligands in the \ce{WO6} octahedron and reorientation of a coordinated water molecule costs too much energy.  Second, the nearest-neighbor distance between two O\textsubscript{i} ions is 3.55~{\AA}, while the formation of \ce{H5O2+} usually requires much closer O-O distances on the order of 2.6$\sim$2.8~{\AA}. \cite{Muguet1996173}

Instead of the Grotthuss mechanism, proton diffusion in \ce{WO3.2H2O} proceeds through the octahedral tungsten oxide layer, shaded green in Figures~\ref{fig:dihydrate}a and \ref{fig:diffusion}a. The proton diffusion path is shown red in Fig.~\ref{fig:diffusion}a and the relevant O\textsubscript{b} ions within the \ce{WO6} layer are labelled O(1) through O(7). The diffusion mechanism consists of two steps. In the first step, the proton starts out at a site near a bridging O\textsubscript{b} ion shared by two \ce{WO6} octahedra, such as near the O(1) ion in the upper-left corner of Fig.~\ref{fig:diffusion}a. Then the proton rotates approximately 180 degrees around the W-O\textsubscript{b}-W axis [W-O(1)-W in Fig.~\ref{fig:diffusion}a]. This rotation is facilitated by the fact that the proton remains bonded to one of the bridging oxygen's nonbonding $2p$ orbitals, which rotates with the proton, as can be seen in the atom- and momentum-decomposed eDOS curves (see Fig. S5 in ESI\dag). To accommodate the proton in the final state, there are pronounced rotations of the surrounding octahedra around the $b$ axis, which widen the angle O(3)-O(1)-O(2) and create a local environment for the proton that is similar to the initial state (see Movie S1 in ESI\dag). To examine if the strain caused by these octahedral rotations can be relaxed in a bulk solid, a larger supercell containing 32 formula units was used to verify the results (see Movie S2 in ESI\dag). The calculated activation barrier for proton rotation in the larger supercell is 0.21 eV, which agrees well with the values obtained from the smaller 16 formula unit supercell. It also shows that the corner-linked octahedral layer is quite flexible with respect to correlated rotations of the octahedra.

In the second step, the proton hops towards the diagonal bridging oxygen site, e.g., from O(1) to O(4) or from O(4) to O(7) in Fig.~\ref{fig:diffusion}a. The small difference in the energy barriers for the O(1)$\rightarrow$O(4) and O(4)$\rightarrow$O(7) hops, shown in Fig.~\ref{fig:diffusion}b, is caused by the different arrangement of coordinated water molecules in the octahedra and by the intrinsic distortions of the octahedra.  After the second step, the proton rotates again as in the first step. The total activation energy for this two-step process is 0.42 eV, which is in good agreement with the value of 0.36 eV deduced from experimental measurements.\cite{li2000proton}

Hopping between neighboring oxygen sites in the (010) plane [for instance, from O(4) to O(5) or from O(4) to O(6)] has a significantly higher energy than the two-step process outlined above because the corresponding transition state involves large distortions of the \ce{WO6} octahedra (see Movie S3 in ESI\dag). Hence, protons prefer to move diagonally as plotted in Fig.~\ref{fig:diffusion}b. Although the two-step diffusion process represents a one-dimensional diffusion pathway, the diffusion as a whole is two-dimensional due to the symmetry of the \ce{WO6} layer. Indeed, the described mechanism also works for proton diffusion in the perpendicular O(5)$\rightarrow$O(6) direction.

\subsubsection{$\mathbf{\gamma}$-WO\textsubscript{3}. }

Since a proton diffuses through the octahedral layer in the dihydrate without direct participation of the structural water molecules, it is reasonable to expect that a similar proton diffusion mechanism should operate in other structurally related tungsten oxides, in particular $\gamma$-\ce{WO3}. As expected, we find that the proton diffusion in $\gamma$-\ce{WO3} also follows the same two-step process as that in the dihydrate (see Fig.~\ref{fig:monohydrate}c). The barrier for the proton rotation step indicated by the dashed black arrow in Fig.~\ref{fig:monohydrate}c is 0.12 eV, while the hop (labelled by the solid blue arrow) has an activation energy of 0.35 eV. This process results in an activation energy of 0.35 eV for proton diffusion in $\gamma$-\ce{WO3}, in good agreement with the value of 0.4 eV measured by  Randin {\it et al.}\cite{randin1982proton} The predicted proton insertion voltage in $\gamma$-\ce{WO3} is 0.11 V, which reproduces the redox peak position (0.2 V relative to SHE) in published cyclic voltammetry measurements with slow sweep rates.\cite{di1978electrochromism}

In what follows we hypothesize how structural and compositional modifications may affect proton transport rates in $\gamma$-\ce{WO3}. It has been reported that heterogeneous atom doping, such as nitrogen doping, can narrow the band gap of tungsten oxide and improve the absorption of sunlight in water splitting applications.~\cite{N2dopingWO3} Effect of heterogeneous doping on proton diffusion in $\gamma$-\ce{WO3} is difficult to predict in general, but it is conceivable that the associated disorder may stiffen the corner-linked octahedral network and therefore hinder concerted atomic relaxations during the proton rotation and hopping steps. This would lead to slower diffusion rates. Faster kinetics of proton diffusion was observed in nanoparticle \ce{WO3},~\cite{ADMA200501953} and we suggest that it might be contributed by the surface proton diffusion. Finally, we note that the mechanisms of lithium and sodium diffusion in $\gamma$-\ce{WO3} will likely be different from that of the proton diffusion.  Due to the nature of ionic bonding of alkaline ions in $\gamma$-\ce{WO3} rather than covalent bonding of H-O, the \ce{Li+}/ \ce{Na+} ion occupies the site in the center of the cage formed by eight \ce{WO6} octahedra.~\cite{PhysRevB.54.2436} Activation energy will be determined by hopping between these cages, not by the proton transfer steps shown in Figure~\ref{fig:monohydrate}c.

\subsubsection{WO\textsubscript{3}$\cdot$H\textsubscript{2}O. }

In the dilute limit (\ce{H$_{y}$WO3.H2O}, $y=0.06$), the calculations predict that the proton adsorption energy at the O\textsubscript{t} site is 0.25 eV lower than at the O\textsubscript{b} site. In the absence of structural water, the O\textsubscript{t}-H dimer points to the nearest O\textsubscript{t}, i.e., O\textsubscript{t1}-H points to O\textsubscript{t2} in Fig.~\ref{fig:monohydrate}d. A zigzag chain formed by the O\textsubscript{t} ions along the [100] direction acts as a fast proton diffusion channel with an activation energy of only 0.07 eV, much lower than the calculated activation energies in $\gamma$-\ce{WO3} and \ce{WO3.2H2O}.  Even though this value suggests that transport of individual protons in defect-free monohydrate crystals can be very fast, we will argue below that the presence of crystalline defects and strong repulsive interactions with other protons are likely to slow the diffusion significantly due to the one-dimensional fast diffusion pathways.

We first discuss the energetics of proton adsorption. Using $10.8\times10.7\times10.4$~{\AA} supercells with 16 formula units, the calculated dilute limit voltages of proton intercalation at the O\textsubscript{t} and O\textsubscript{b} sites are 1.18 V and 0.93 V, respectively.  These numbers are much higher than those in $\gamma$-\ce{WO3} and \ce{WO3.H2O}, showing that proton intercalation is energetically highly favorable and suggesting that the dilute limit is not relevant to the conditions used in experimental kinetic measurements.
Adding an additional proton to the supercell (corresponding to a concentration $y = 0.13$ in H$_y$\ce{WO3.H2O}) changes the energetics dramatically:  the second proton prefers to adsorb at the bridging oxygen site and the lowest energy configuration corresponds to an intercalation voltage of 0.11 V. This calculation shows that the mutual repulsion of protons is very strong, most likely due to poor screening of the positive charges of the intercalated protons. The calculated diagonal entries of the high-frequency dielectric tensor $\epsilon^\infty_{\alpha\beta}$  are 5.3, 3.3, and 5.5, lending support to this idea. The diagonal entries of the static dielectric tensor $\epsilon^0_{\alpha\beta}$ are calculated to be 182.0, 4.6, and 50.9. The high values for the in-plane screening are reflective of the easy deformability of the \ce{WO6} octahedral layer, while in the direction perpendicular to the layers the screening remains small. While the values of the dielectric constants are suggestive, the most persuasive evidence for strong proton-proton repulsion comes for the pronounced concentration dependence of the calculated intercalation voltages.

The atomic configuration of the O\textsubscript{b}-H dimer is practically the same as in the dihydrate (see Fig.~\ref{fig:protonE}). The activation energy for proton diffusion within the octahedral layers is found to be 0.52 eV, significantly higher than the value calculated for the zigzag chain, but comparable with the activation energies in $\gamma$-\ce{WO3} and \ce{WO3.H2O}. The calculated migration barrier between the O\textsubscript{t} and O\textsubscript{b} sites is very high (0.95 eV) and hence difficult to overcome at ambient conditions. This indicates that the O\textsubscript{t}-O\textsubscript{t} and O\textsubscript{b}-O\textsubscript{b} diffusion paths are isolated and protons cannot make a transition from one to the other.

We hypothesize that with increasing proton concentration all of the one-dimensional zig-zag channels will be eventually blocked by one or more protons adsorbed within the octahedral layers at one of the nearby O\textsubscript{b} sites. For the channel to re-open, these bridge-site protons have to diffuse {\it within\/} the layer, which requires the two-step process shown in Fig.~\ref{fig:diffusion}. A key element in the first (O-H rotation) step of this mechanism is a concerted rotation of the surrounding \ce{WO6} octahedra around the $b$ axis to make space for the proton in the final state. While it is relatively inexpensive in the dilute limit, the presence of additional protons will hinder the rotation because it would not allow compressing the O\textsubscript{b}-O\textsubscript{b}-O\textsubscript{b} angle at their adsorption sites. We have attempted to quantify the energetic cost of such an event, but unfortunately have not been able to obtain a converged transition state structure using the NEB method. Systematic studies of proton diffusion at high concentrations would be interesting, but are beyond the scope of this paper.

To summarize, we propose that the mere presence of fast diffusion pathways in the monohydrate does not guarantee good proton transport properties. Due to the much stronger preference for proton uptake and due to weak screening of Coulomb interactions, repulsive interactions between the intercalated protons are likely to block the fast one-dimensional diffusion channels as well as inhibit the two-dimensional diffusion observed in $\gamma$-\ce{WO3} and \ce{WO3.H2O}. Presence of structural defects, which are thermodynamically favored in sufficiently large samples, is likely to further degrade the diffusion kinetics through the one-dimensional zig-zag channels in a manner similar to the well-studied case of olivine battery materials. \cite{Olivine}

\section{Conclusions}

In summary, crystal structures, electronic properties, hydrogen intercalation energetics, and migration barriers in the \ce{WO3.$x$H2O} ($x=0,1,\text{ and }2$) family of hydrated oxides have been investigated using DFT calculations. In contrast to the na\"ive expectation that proton diffusion in hydrates should occur via the Grotthuss mechanism, our results show that the intercalated proton diffuses through the layers of corner-sharing \ce{WO6} octahedra and does not involve the hydrogen-bonded \ce{H2O} network in \ce{WO3.2H2O}. There are several factors that combine to produce this interesting behavior: (1) intercalated protons prefer to bind to one of the bridging oxygen sites due to the existence of non-bonded $2p$ orbitals,  (2) a hydronium ion in the water layer is thermodynamically unstable, and (3) the distortion of the octahedral \ce{WO6} layer introduced by the adsorbed proton can be relaxed in a small region, resulting in lower proton migration barriers. The same diffusion mechanism and similar proton migration barriers have been found in the room-temperature $\gamma$-phase of \ce{WO3}. In contrast, the monohydrate \ce{WO3.H2O} follows different mechanisms of proton diffusion. In the dilute limit, the intercalated proton diffuses along one-dimensional zig-zag pathways running in the [100] direction between the octahedral layers. The zig-zag pathways are formed by apical O\textsubscript{t} ions and the calculated migration barrier is only 0.07 eV. Due to the stronger energetic preference for proton absorption, a much higher concentration of protons is predicted to exist in H$_y$\ce{WO3.H2O} under typical experimental conditions. At non-dilute concentrations, our results indicate that strong repulsive interactions exist between the intercalated protons. We hypothesize that proton-proton repulsion in combination with structural defects will block the fast one-dimensional diffusion channels, while the presence of a high concentration of protons in the octahedral layers will hinder the rotation step necessary for the two-dimensional intralayer mechanism operating in $\gamma$-\ce{WO3} and \ce{WO3.2H2O}. This argument provides insight into the reasons why bulk proton transport in \ce{WO3.H2O} is suppressed, as indicated by experimental measurements which can only detect a surface contribution.\cite{li2000proton}

\section*{Acknowledgements}

This work was supported as part of the Molecularly Engineered Energy Materials, an Energy Frontier Research Center funded by the US Department of Energy, Office of Science, Basic Energy Sciences under Award Number DE-SC0001342. Calculations were performed using resources of the National Energy Research Scientific Computing Center (NERSC), which is supported by the Office of Science of the U.S. Department of Energy under Contract No. DE-AC02-05CH11231. F.Z. is supported at Lawrence Livermore National Laboratory under D.O.E. Contract No. DE-AC52 -07NA27344.




\footnotesize{
\bibliography{ref} 
\bibliographystyle{rsc} 
}

\end{document}